\documentclass{aa}  

\usepackage{graphicx}
\usepackage{txfonts}
\usepackage[]{hyperref}
\begin{document}

   \title{Hot Jupiters, cold kinematics}

   \subtitle{High phase space densities of host stars reflect an age bias}

   \author{Alexander J. Mustill
          \inst{1}
          \and
          Michiel Lambrechts
          \inst{1}
          \and
          Melvyn B.~Davies
          \inst{1,2}
          }

   \institute{Lund Observatory, Department of Astronomy and Theoretical 
     Physics, Lund University, Box 43, 
              221\,00 Lund, Sweden\\
              \email{alex@astro.lu.se}
              \and
              Centre for Mathematical Sciences, Lund University, 
              Box 118, 221\,00 Lund, Sweden
             }

   \date{Received XXX; accepted YYY}

 
  \abstract
   {The birth environments of planetary systems 
     are thought to influence planet formation and 
     orbital evolution, through external photoevaporation 
     and stellar flybys. Recent work has 
     claimed observational support for this,
     in the form of a correlation between 
     the properties of planetary systems and 
     the local Galactic phase space density 
     of the host star. In particular, 
     Hot Jupiters are found overwhelmingly around
     stars in regions of high phase space density, 
     which may reflect a 
     formation environment with high stellar density.}
   {We instead investigate whether the high phase space density 
     may have a galactic kinematic origin:  
     Hot Jupiter hosts may be biased
     towards being young and therefore kinematically cold, 
     because tidal inspiral leads to the 
     destruction of the planets on Gyr timescales, 
     and the velocity 
     dispersion of stars in the Galaxy increases 
     on similar timescales.}
   {We use 6D positions and kinematics from \emph{Gaia}
     for the Hot Jupiter hosts and their neighbours, 
     and construct distributions of the phase space density.
     We investigate correlations between 
     the stars' local phase space density and 
     peculiar velocity.}
   {We find a strong anticorrelation between the 
     phase space density and the host star’s
     peculiar velocity with respect to the Local Standard of Rest. 
     Therefore, most stars in ``high-density''
     regions are kinematically cold, which may be 
     caused by the aforementioned bias towards detecting Hot Jupiters 
     around young stars before the planets' tidal destruction.}
   {We do not find evidence in the data for Hot Jupiter hosts 
     preferentially being in phase space overdensities 
     compared to other stars of similar kinematics,
     nor therefore for their 
     originating in birth environments of high stellar density.}


   \maketitle
%

\section{Introduction}

The birth environment of a planetary system---the 
size and density of the stellar cluster or association 
where the system forms---is thought to have an impact on the nature 
of the planetary system. This can occur through external 
photoevaporation of the protoplanetary disc, and/or through
stellar flybys truncating the disc or perturbing 
formed systems \citep{delaFuenteMarcos97,LaughlinAdams98,
  Adams+06,Malmberg+07,Winter+18,Li+19,Li+20b,Li+20a}. Direct evidence
of the impact of birth environments on planetary system formation 
is hard to obtain, because of the low number (in absolute terms) of 
planets found in clusters compared to the field, 
and the challenges of detecting planets around 
young stars. \citet[henceforth W20]{Winter+20} 
recently proposed an ingenious way around this, 
by using the local phase space density---the 
density of nearby stars in the 6D phase space 
of Galactic position and velocity---of an 
exoplanet host star as a proxy for the crowdedness of 
its birth environment. By assuming that the current 
density reflects the past density at birth, 
W20 could look for correlations 
between this density and the properties of 
planetary systems.

One of the most significant results from W20 
was that the host stars of Hot Jupiters are 
nearly always in ``high-density'' regions of phase space.
This is naturally explained if the primary migration
channel for Hot Jupiters is dynamical excitation through
planet--planet scattering and/or Lidov--Kozai cycles,
followed by tidal dissipation 
\citep{RasioFord96,WeidenschillingMarzari96,WuMurray03,FabryckyTremaine07}. 
Here, the external
dynamical perturbations in a dense birth environment
would provide the trigger for this high-eccentricity
migration to begin
\citep{MalmbergDC07,ParkerGoodwin09,Malmberg+11,ParkerQuanz12,
  Brucalassi+16,Rodet+21}. Disc migration offers an alternative 
channel to produce Hot Jupiters \citep{Lin+96}, which might 
also be affected by the environment through photoevaporation 
of the protoplanetary disc \citep{Winter+18}.

\begin{figure}
  \includegraphics[width=0.5\textwidth]{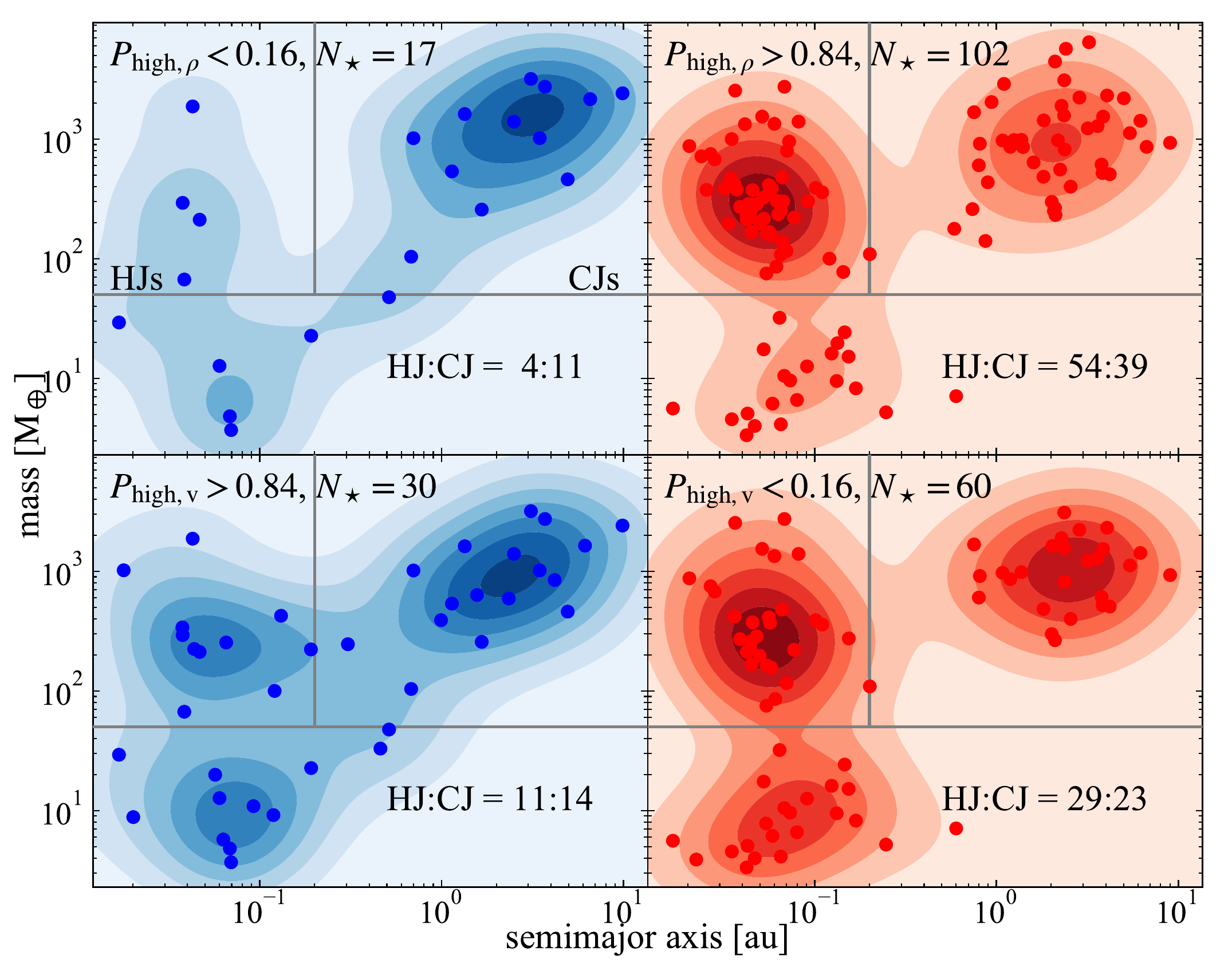}
  \caption{Semimajor axes and masses of planets whose host stars 
  have masses between $0.7$ and $2.0\mathrm{\,M}_\odot$ 
  and ages between $1.0$ and $4.5$\,Gyr. In the top panels they 
  are distinguished by the host stars' phase space density, 
  in the bottom panel by the host stars' peculiar velocity. 
  Typically, a low velocity corresponds to a high phase 
  space density and vice versa. We see an abundance 
  of Hot Jupiters in  the high-density 
  and the low-velocity populations. Note that stars that cannot 
  be unambiguously assigned to one of the 
  populations are not included, and that some stars have 
  multiple planets.}
  \label{fig:a_M}
\end{figure}

This interpretation of W20's finding, though, 
relies on the assumption that a high local phase 
space density for a star at the present time reflects 
a high density of its formation environment. 
Here we examine this assumption. 
In Hamiltonian mechanics, Liouville's Theorem indeed 
states that phase space density 
is constant along trajectories, implying it would be inherited 
from a star's formation site. But this does not apply to stars 
in the Galaxy on Gyr timescales, because
the Galactic potential is not time-independent, and the dynamics are
not completely collisionless, violating the conditions 
for Liouville's Theorem to apply. In particular, stellar populations get
``heated'' with age and increase their velocity dispersion and vertical 
scale height, through interactions with giant molecular clouds 
and spiral arms \citep{SpitzerSchwarzschild51,Wielen77,
  DeSimone+04,Nordstroem+04}. Numerical simulations
\citep{Kamdar+19a,Kamdar+19b} find that the imprint of a birth
cluster, in comoving conatal pairs and phase space
overdensities, largely disappear after $\sim1$\,Gyr.
Hence, it is not clear that the foundational assumption of
W20 holds. Instead it is possible that stars' phase space
density reflects coarser features of Galactic structure and kinematics,
such as disc heating with age, or the existence of a Galactic 
thick disc of larger scale height than the thin disc 
to which the Sun belongs \citep{GilmoreReid83}, 
rather than the nature of the birth cluster.
While noting that Galactic dynamics could play
a role in determining the phase space density,
W20 interpreted their results in the context of the
birth environment hypothesis.

Here we show that, in
general, the majority of stars which are currently in ``high density'' 
regions of phase space as defined by W20 simply have cold kinematics 
in the Galactic disc: near-circular orbits and little 
vertical motion. The ``high density'' classification thus 
relates to the lower average age and 
lesser kinematic heating of the stars, and not to 
a memory of their birth environment.

\section{Methods: Mahalanobis distance and description
  of the phase space density distribution}

We follow W20 in using the Mahalanobis distance \citep{Mahalanobis36}
in 6D phase space to construct the phase space density. 
This is essentially a reoriented and stretched Euclidean 
metric, represented by the quadratic form
\begin{equation}
  d_\mathrm{M}\left(\mathbf{x_1},\mathbf{x_2};C\right) 
  = \sqrt{(\mathbf{x_1}-\mathbf{x_2})C^{-1}
    (\mathbf{x_1}-\mathbf{x_2})^{\mathrm{T}}},
\end{equation}
where $\mathbf{x_1}$ and $\mathbf{x_2}$ are two 
(6D) phase space positions and $C$ is the $6\times6$ 
covariance matrix of the whole sample
\begin{equation}
C_{ij} = \big\langle \big(x_i-\langle x_i\rangle\big)
\big(x_j-\langle x_j\rangle\big)\big\rangle
\end{equation} 
for $i,j=1\ldots6$, where $\langle \cdots \rangle$ denotes 
the mean.
It has the advantage of defining a distance
in a space whose dimensions have different dynamic 
ranges and different physical dimensions. However, 
as it returns a rescaled, dimensionless quantity, 
its physical interpretation is not obvious. 
We will therefore relate the density derived from 
this metric to the host stars' corresponding 
physical quantities, especially the peculiar velocity:
the star's velocity with respect to that of a circular 
orbit in the Galactic plane (the Local Standard of Rest).

We begin by following W20 
as closely as possible to ensure a direct comparison with their work. 
For each target star we:
\begin{itemize}
\item Query all objects in \emph{Gaia} Data Release~2 (DR2)
  or Early Data Release~3 (EDR3) 
  \citep{Gaia2016,Gaia2018,Gaia2020} within
  80\,pc of the target. The criterion for inclusion is that 
  the star possess a radial velocity measured by \emph{Gaia} 
  as well as a positive parallax.
\item Convert the astrometric, positional and RV information 
  from \emph{Gaia} to a heliocentric Cartesian position and velocity. 
  The distance was obtained by inverting the parallax. A velocity correction 
  from the heliocentric rest frame to the local standard of rest 
  (the velocity of a body on a circular orbit at the Sun's 
  position in the Galaxy) from \cite{Schoenrich+10} was then applied: 
  $(U,V,W)_\odot = (11.1, 12.24, 7.25)\mathrm{\,km\,s}^{-1}$.
\item Define the Mahalanobis metric on the sample, 
  using the covariance matrix of the positions and 
  velocities of all stars within 80\,pc of the target.
\item For stars with at least 400 neighbours within 40\,pc,
  randomly choose up to 600 such neighbours. 
  For each of these, as well as the target, 
  find the 20th nearest neighbour by the Mahalanobis distance 
  $d_\mathrm{M,20}$, 
  and use this to define the local 6D phase space density 
  $\rho_{20} = 20d_\mathrm{M,20}^{-6}$.
\item Normalise $\rho_{20}$ so that the median of the 
  distribution is 1.
\item Fit a two-component Gaussian mixture model to 
  the distribution of the logarithm of the rescaled density. 
  Outliers greater than two standard deviations 
  from the mean, or with densities $\rho_{20}>50$, are clipped 
  before fitting the model.
\item Remove systems where a one-component model is a 
  good fit to the density distribution ($p>0.05$ on a KS test); 
  three systems are so removed.
\item Calculate the probability that the target star 
  was drawn from the high-density or the low-density component of 
  the Gaussian mixture model. If $\rho>50$ assign it to the 
  high density population.
\end{itemize}
After this, W20 analysed differences between the ``high-density'' 
population ($P_\mathrm{high}> 0.84$) and the ``low-density'' 
population ($P_\mathrm{high}<0.16$). The power of this approach 
is illustrated in Figure~\ref{fig:a_M}, where we show the semimajor
axes and eccentricities of known exoplanets\footnote{From 
  \url{https://exoplanetarchive.ipac.caltech.edu/index.html}, accessed 
2021-03-11.} whose hosts were 
cross-matched to \emph{Gaia}~EDR3 and whose masses and ages 
are in the range $0.7-2.0\mathrm{\,M}_\odot$ and $1.0-4.5$\,Gyr 
(as in W20). As did W20, we see noticeable differences between the 
distributions of planets orbiting ``high-density'' and ``low-density'' hosts. In 
this paper we focus on the overabundance of Hot Jupiters 
orbiting the ``high-density'' hosts: the ratio of 
Hot to Cold Jupiter hosts is $1.4$ for the high-density hosts and 
only $0.4$ for the low-density hosts (in this paper,
following W20, we 
define Hot Jupiters as planets with mass $M>50\mathrm{\,M}_\oplus$ and 
semimajor axis $a<0.2$\,au, and Cold Jupiters as planets 
with mass $M>50\mathrm{\,M}_\oplus$ and
semimajor axis $a>0.2$\,au). As we later argue that
tidal effects are likely responsible for the difference
between Hot and Cold Jupiters, we also choose a cut between
Hot and Cold Jupiters of $0.1\,$au. This yields similar ratios
(for high-density hosts, the Hot to Cold Jupiter ratio is
1.2; for low-density hosts, it is again 0.4).
However, 
in the bottom panels of Figure~\ref{fig:a_M} we show the 
same sample of planets but with hosts broken down 
by membership into a high- or low-density component 
of the distribution of peculiar velocities 
$|\mathbf{v}|$ relative to the Local Standard of 
Rest, using the same Gaussian Mixture 
procedure as we used for the densities. Although the difference 
between the planet populations is not so pronounced as when the 
hosts are broken down by phase space density, the same trends 
are seen. This suggests that the stars' peculiar motions 
are in fact conveying most of the information. 
With this in mind, we now step back and 
investigate what the non-dimensionalised, 
rescaled phase space density is physically 
telling us.

\begin{figure*}
  \centering
  \includegraphics[width=0.66\textwidth]{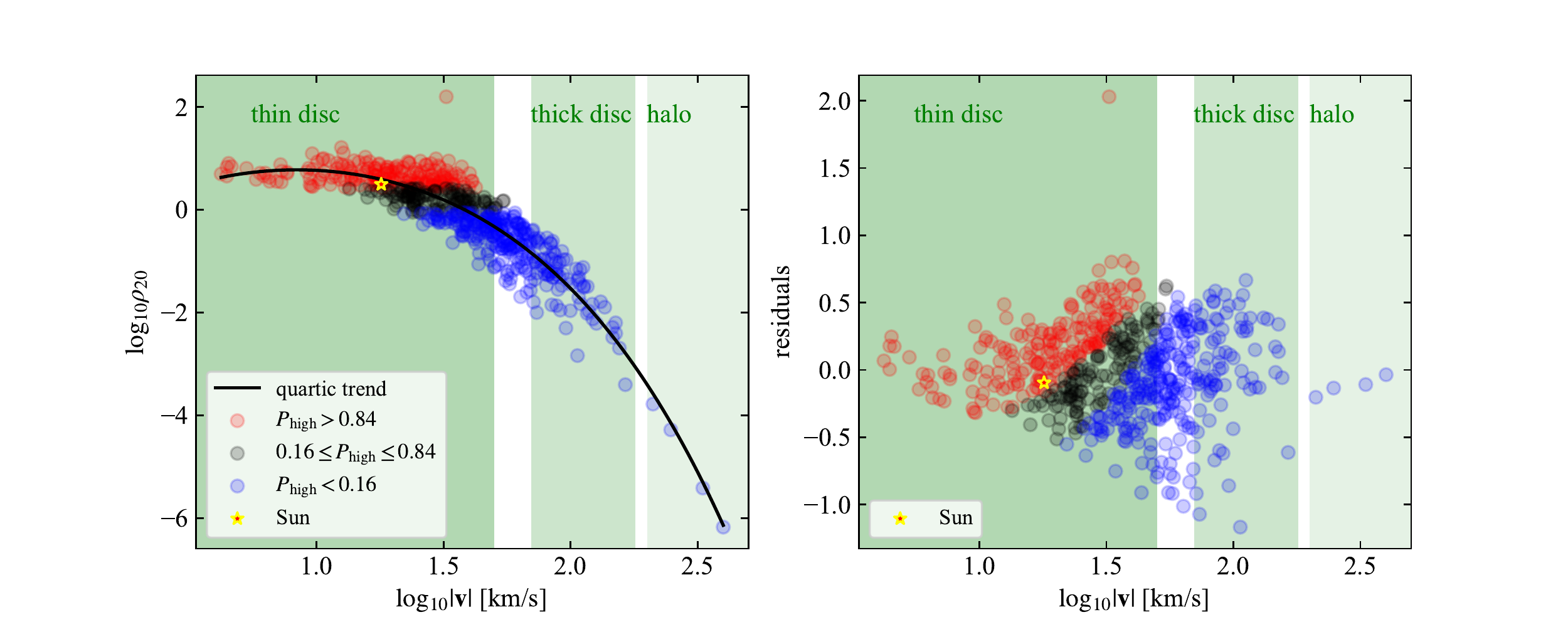}
  \includegraphics[width=0.33\textwidth]{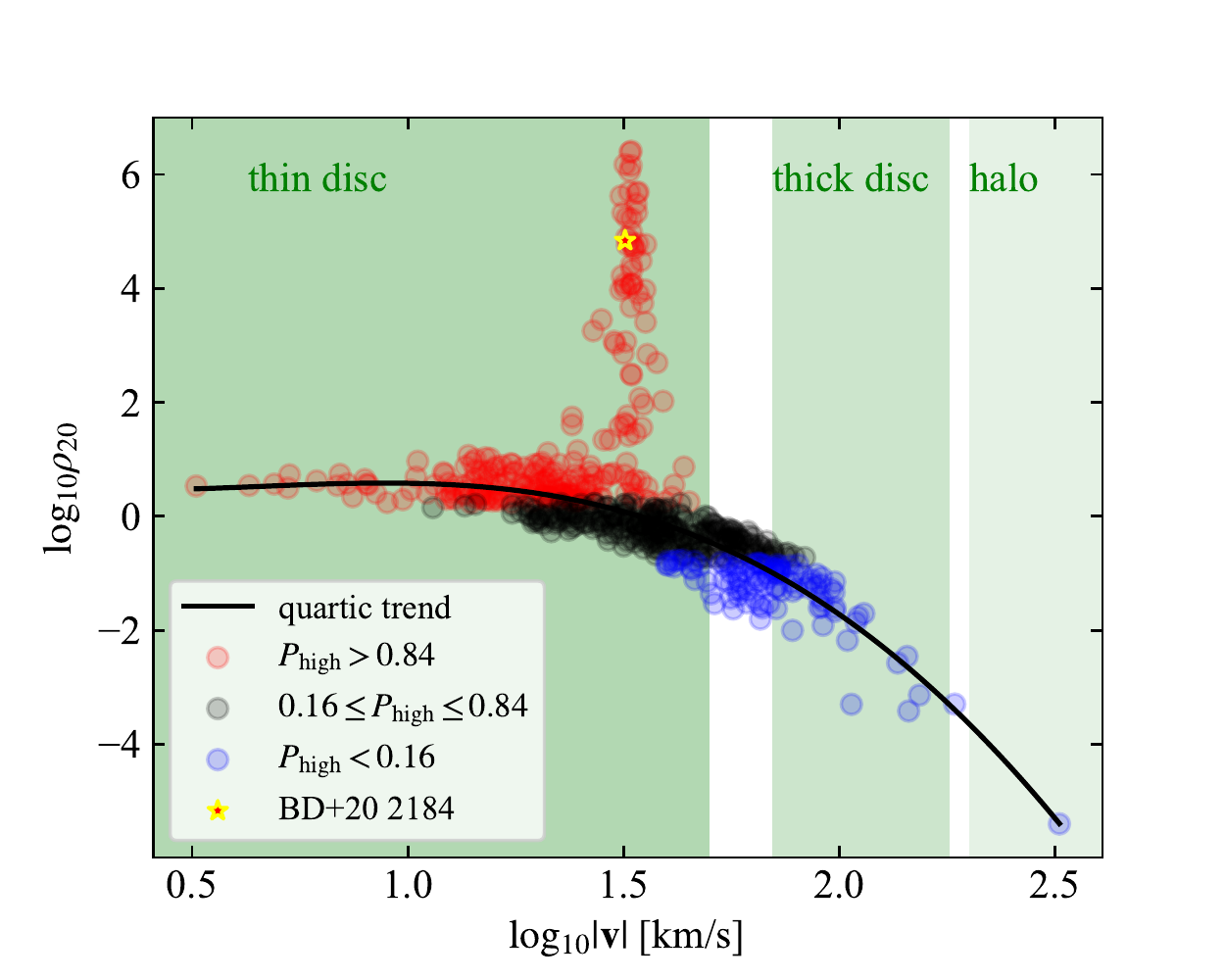}
  \caption{Left: Local 20$^\mathrm{th}$-nearest neighbour
    phase space density, and probability
    of membership in the high-density population,
    for 600 stars within 40\,pc of the Sun. 
    We show the fitted quartic trend of the density 
    as a function of peculiar velocity. 
    The outlier at $\rho\sim150$ is \emph{Gaia}~EDR3~3277270538903180160
    (LP~533-57, HIP~17766), a Hyades member.
    Centre: The residuals to the fit, after the trend 
    is removed. Right: The densities and 
    peculiar velocities of 600 neighbours 
    of Pr0201 (BD+20~2184), a Hot Jupiter host 
    in the Pr{\ae}sepe cluster. Pr0201 and 
    other Pr{\ae}sepe members stand out 
    above the trend.
    Green background shading shading shows 
    the approximate bounds for membership in the Galactic thin 
    disc, thick disc and halo, from \cite{Bensby+14}.}
  \label{fig:V_rho}
\end{figure*}

\section{Phase space density and Galactic velocities}

We begin by taking the Sun as a case study. W20 identified the Sun 
as belonging to a phase-space overdensity. This may reflect the 
Sun's origin in a reasonably large, dense cluster 
\citep{Adams10,Pfalzner+15}. 
However, we note here that the Sun has a rather low 
peculiar motion for its age \citep{Wielen+96,Gonzalez99}. 
The colour--magnitude diagram for our sample 
of Solar neighbours from \emph{Gaia} EDR3 is shown in Figure~\ref{fig:CMD}. 

From the Solar neighbours within 40\,pc, we have randomly selected 
600 and calculated their local phase space density as described 
above. We show histograms of the phase space density distribution 
in Figure~\ref{fig:GMM}. In common with the distributions for the 
neighbourhoods of other target stars, the distribution 
is poorly fit by a single lognormal. Instead, there is typically a 
steep cutoff at high density, plus a few high-density outliers 
often  associated with known clusters or moving groups, and 
a shallower tail towards lower densities. A two- or even higher-component 
fit is usually superior; in fact, three or more components 
are often favoured by the Aikake and/or Bayesian 
Information Criteria, suggesting that 
the distribution is rather a continuous spectrum than the sum 
of a high-density and a low-density lognormal. 
The middle panel of 
Figure~\ref{fig:GMM} shows the decomposition into two components. 
The Sun lies near the peak of the high-density component. 
Finally, the right panel shows the probability that the Sun belongs to the 
high-density component, which we calculate to be $0.88$.
The probability of belonging to 
the high-density component actually decreases slightly at high densities 
on account of the breadth of the low-density component. Visual inspection of 
several distributions showed that this is usually not too extreme a problem; 
it was helped by clipping the outliers before fitting the Gaussian mixture 
model as described above. The Sun is found to have a high 
probability of belonging to the high-density population, as W20 found.
In Figure~\ref{fig:GMM_v} we show the equivalent Gaussian Mixture models 
for the velocity distribution; here the Sun belongs to the low-velocity component
(the probability it belongs to the high-velocity component is $0.03$.)

As the Mahalanobis density is constructed from 
both spatial and kinematic information, we now ask 
which of these is most significant. We begin with 
the velocities, as suggested in Figure~\ref{fig:a_M}.
Figure~\ref{fig:V_rho} shows the
local phase space density for neighbours of the
Sun and BD+20\,2184 (alias Pr0201, a Hot Jupiter host
in the Pr{\ae}sepe open cluster, \citealt{Quinn+12}), as 
a function of the stars' peculiar velocities.
In each case we see a strong correlation:
the phase space density is primarily determined by a star's
peculiar velocity. ``High-density'' stars are those with low
peculiar velocities, ``low-density'' stars those with high
peculiar velocities. We show in Figure~\ref{fig:V_rho} with
the background colouring the approximate velocity ranges
corresponding to membership in the Galactic thin
disc, thick disc and halo \citep{Bensby+14}. The ``low-density''
stars appear to be a heterogeneous mix of dynamically hot thin disc
stars, thick disc stars, and a handful of halo stars, while the
``high-density'' stars are lower velocity thin disc stars. A natural
interpretation, then is that stars move from the high-density population
to the low-density population as they age and are kinematically
heated in the disc.

We also see in
Figure~\ref{fig:V_rho} the Pr{\ae}sepe cluster at 
$|\mathbf{v}|\approx30 - 40\mathrm{\,km\,s}^{-1}$, 
standing out from the field star trend.
The field star population is rather smooth;
we detrend this in the next section.

In contrast, large-scale spatial structure (\emph{i.e.,} 
10s of pc) has little 
influence on the phase space density. Figure~\ref{fig:D_rho} shows the 
local phase space density for neighbours of the 
Sun and BD+20\,2184. For 
the Sun, we see little large-scale spatial structure: the 
Mahalanobis phase space density of a star is not strongly dependent on 
its distance to the Sun. For the neighbours of BD+20\,2184, 
the other Pr{\ae}sepe members are clearly identified 
by the Mahalanobis density measure as a distinct group, 
attaining densities of $\gtrsim10^3$ 
within around 10\,pc of the target. However,
most of the ``high-density'' stars as defined by the 
Gaussian Mixture model are not cluster 
members: 167 ``high-density'' stars lie beyond 20\,pc 
of BD+20~2184, and only 74 within 20\,pc, those beyond 
20\,pc having furthermore densities only a little 
higher than the rest of the field star population, rather
than orders of magnitude higher as is the case for the 
cluster members.

\section{Hunting for a trend in the residuals}

\label{sec:residuals}

\begin{figure*}
  \centering
  \includegraphics[width=0.49\textwidth]{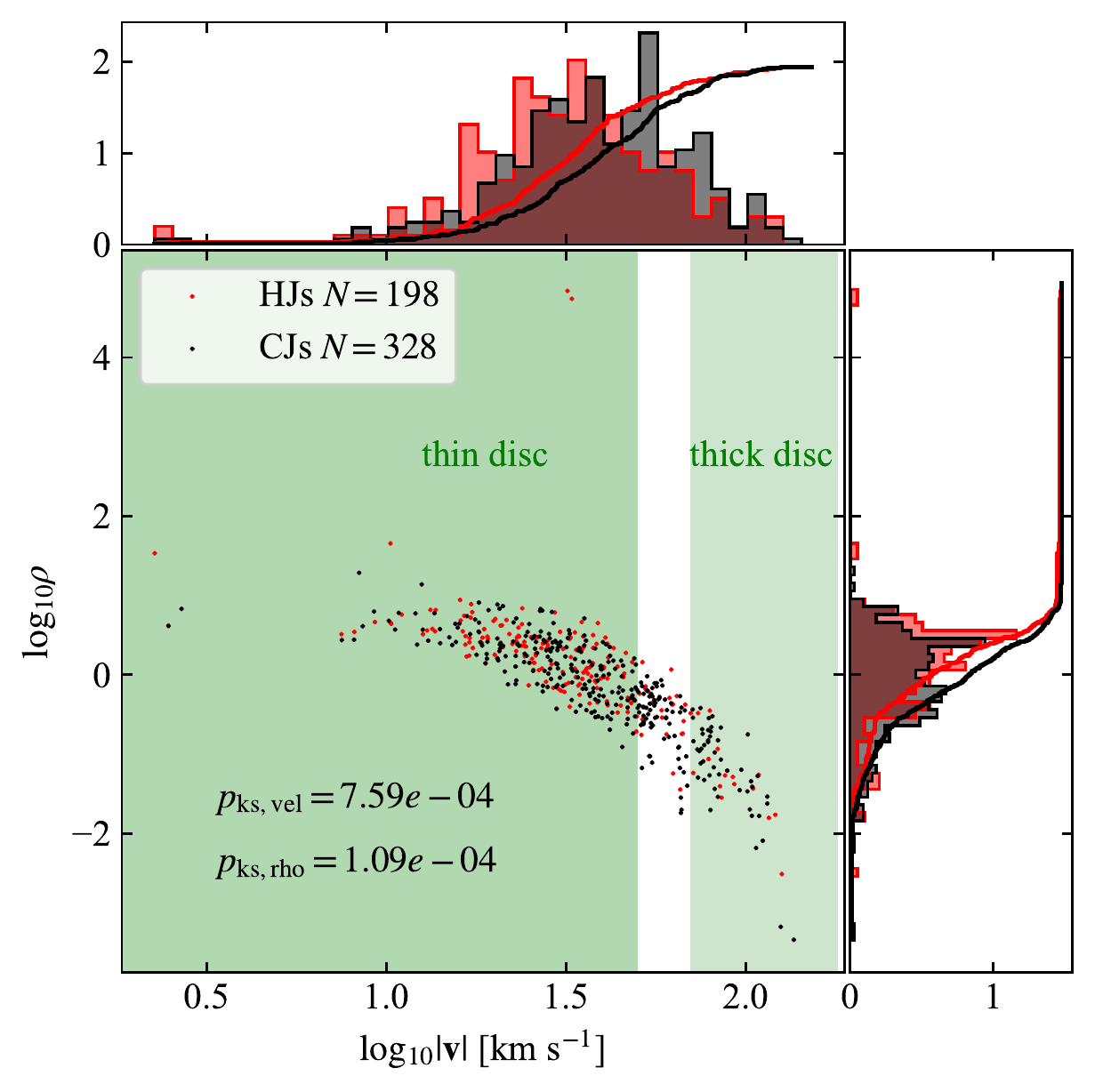}
  \includegraphics[width=0.49\textwidth]{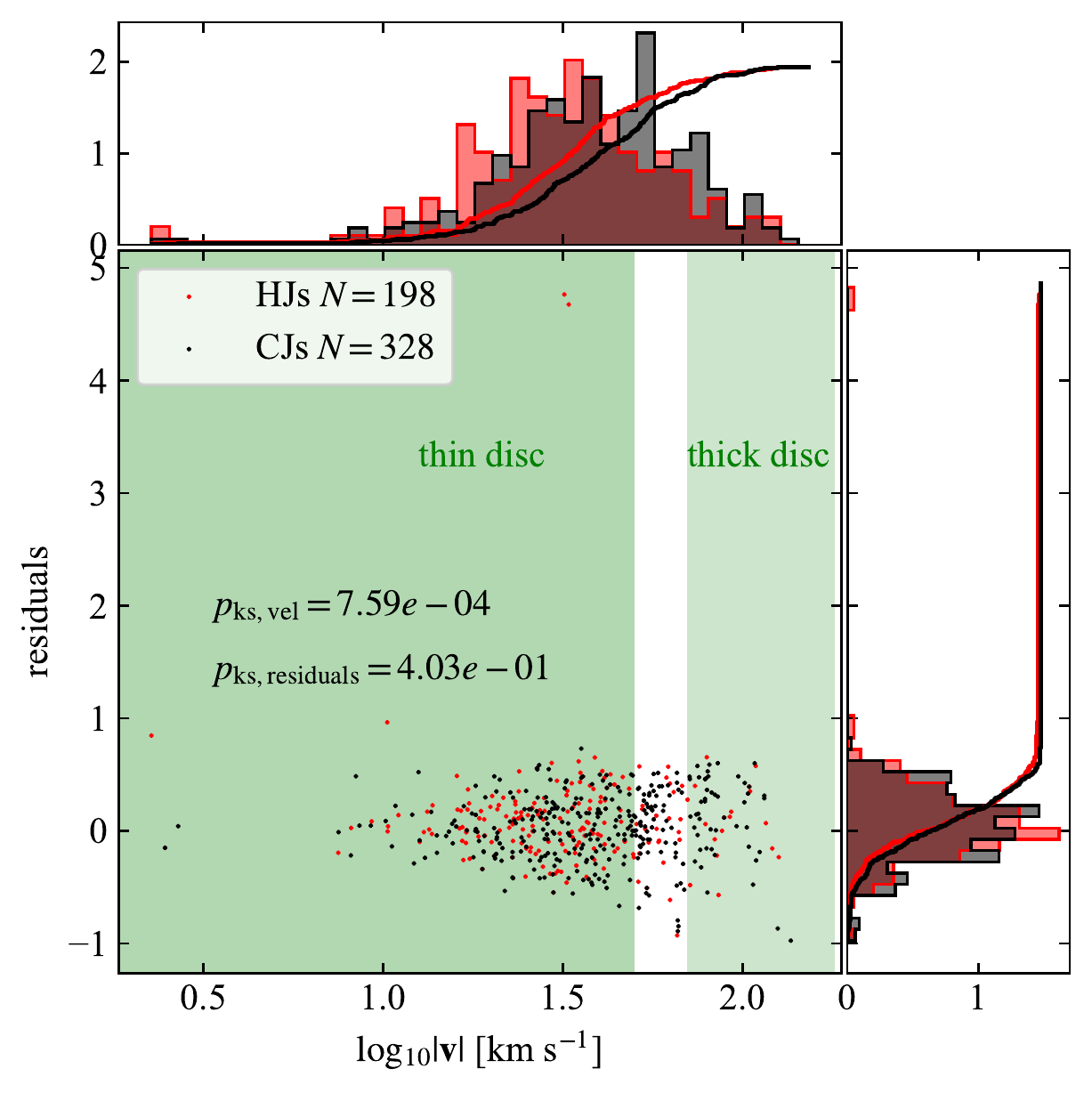}
  \caption{Left: phase space density versus peculiar velocity
    for hot Jupiter host stars (`HJs') and for cold Jupiter host
    stars (`CJs'). Both marginal distributions (top and right 
    sub-panels) are statistically
    significantly different: the KS-test $p$-values are shown in the
    figure. HJ hosts have lower velocity and higher density than 
    CJ hosts. Right: residuals to the detrended phase space density
    versus peculiar velocity for the same stars. The difference in
    the distribution of residuals between the Hot and Cold Jupiter
    populations is not significant ($p_\mathrm{KS,residuals}=0.40$).}
  \label{fig:HJs_CJs}
\end{figure*}

We now investigate whether there is any correlation of phase 
space density with the presence of a Hot Jupiter after correcting for 
the dependence of phase space density on peculiar velocity. 
Are Hot Jupiter hosts ``high-density'' when compared to stars 
of similar kinematics? This could be the case if, for example, 
stars are born from regions of similar $|\mathbf{v}|$ but 
different densities, and this density difference persists 
as the stars get heated in the disc.

First we detrend the $\log\rho-\log|\mathbf{v}|$ relation. For each of 
our host stars, we fit a quartic polynomial to $\log\rho$ as a 
function of $\log|\mathbf{v}|$ for each of its 
sample of neighbours. The example of the
Sun is shown in Figure~\ref{fig:V_rho}. In performing this fit, 
we exclude densities greater than 50 to avoid the fit being 
biased by clusters: we wish to fit only field stars. We see in 
Figure~\ref{fig:V_rho} that, although the Sun is a ``high-density'' host, 
it lies very close to the fitted trend and is quite unremarkable given 
its cold kinematics.

We repeat this detrending for each of a sample 
of Hot Jupiter hosts (planet mass $\ge50\mathrm{\,M_\oplus}$, 
planet semimajor axis $\le0.2$\,au, stellar mass 
$\in [0.7,2.0]\mathrm{\,M}_\odot$\footnote{This is the same 
  definition of Hot Jupiters as that used by W20, except that 
  we have removed the age constraint: when including 
  the age constraint, the sample was too small to see a significant 
  difference in either $\rho$ or $|\mathbf{v}|$.}), 
as well as for a control sample of 
cold Jupiter hosts (same criteria except with 
semimajor axis $>0.2$\,au). We show the phase space densities 
as a function of peculiar velocity in the left-hand panel 
in Figure~\ref{fig:HJs_CJs}; both populations follow a 
similar trend to the Solar neighbours in Figure~\ref{fig:V_rho}
The marginal distributions of both peculiar velocity and phase 
space density differ significantly ($p=7.6\times10^{-4}$ and 
$p=1.1\times10^{-4}$ on KS tests), with the Hot Jupiter hosts having 
lower velocities and higher densities. Our velocity dispersions 
for the Hot and Cold Jupiter hosts are $37.2\mathrm{\,km\,s}^{-1}$
and $43.3\mathrm{\,km\,s}^{-1}$ respectively, 
similar to the values obtained by 
\cite{HamerSchlaufman19}\footnote{Taking the semi-major axis cut
at $a=0.1$\,au, we find $35.3\mathrm{\,km\,s}^{-1}$
and $43.7\mathrm{\,km\,s}^{-1}$ respectively.}. In the right-hand panel, we 
show instead the residuals of each host star to its fitted 
trend. The marginal distributions of these residuals 
are statistically indistinguishable 
($p=0.40$ on a KS test). Thus, after accounting for the 
lower velocity dispersion, there is no evidence that 
the Hot Jupiter hosts are 
located in denser regions of phase space compared 
to the Cold Jupiter control sample.

In Appendix~\ref{sec:compare} we describe an alternative control
experiment, in which we compare the hot Jupiter hosts 
to the 600 randomly-chosen neighbours of the Sun. 
Again, we find that after accounting for the 
dependence of the phase space density on velocity, there is no 
evidence that Hot Jupiter hosts are in regions of high density 
compared to stars with similar kinematics.

\section{Discussion}

\begin{figure}
  \includegraphics[width=0.5\textwidth]{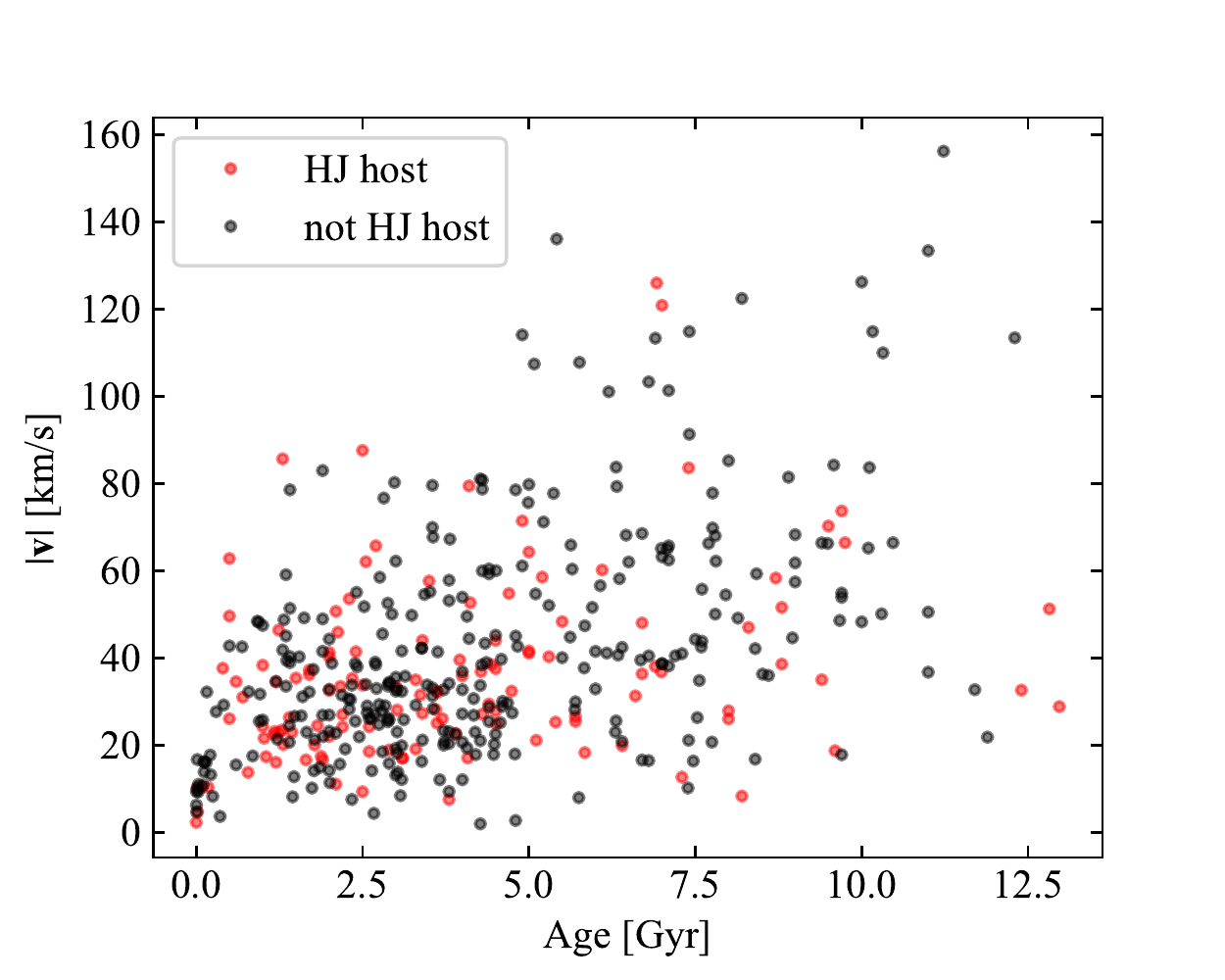}
  \caption{Age--velocity relation for exoplanet host 
    stars with ages given in the NASA Exoplanet Archive. 
    Stars are divided into Hot Jupiter hosts and non-Hot Jupiter 
    hosts. Error bars on the ages are not shown (indeed,
      they are not always available) but can easily be several Gyr.}
  \label{fig:age-vel}
\end{figure}

We have now demonstrated that the main determinant 
of a star's 6D phase space density is the magnitude 
of its peculiar motion, \emph{i.e.,} how much its 
Galactic orbit deviates from a circular orbit 
exactly in the Galactic plane. At the present time, 
and for the past $\sim8$\,Gyr, stars have been typically 
born close to the Galactic midplane with a low peculiar 
velocity, and are heated with age through interactions 
with matter inhomogeneities in the Galaxy. This heating 
occurs on a timescale of Gyr: for eample, with the 
good asteroseismic 
ages derivable from \emph{Kepler} data, \cite{Miglio+21} find 
that  the vertical 
velocity dispersion for thin disc stars 
rises from $\approx10\mathrm{\,km\,s}^{-1}$
at an age of 1\,Gyr to $\approx20\mathrm{\,km\,s}^{-1}$ at
an age of 10\,Gyr. The age--velocity relation for the 
exoplanet host stars with ages given in the 
NASA Exoplanet Archive is shown in Figure~\ref{fig:age-vel}.
We caution that these ages come with large errors, 
as most exoplanet hosts are main-sequence stars, 
and are more over not derived homogeneously 
\citep[see][for discussion on this]{Adibekyan+21}. 
Nonetheless, we do indeed see an increase in the velocity 
dispersion with age.

This age--velocity relation pertains to the Galactic thin 
disc. There also exists a chemically distinct and 
kinematically hotter stellar population, the thick disc, 
whose stars have a higher velocity dispersion 
even than thin disc stars of comparable age \citep{Miglio+21}. 
However, the existence of a clean kinematic separation, 
and the exact relation of kinematics to adundances 
and to the early history of the Galaxy are still debated 
\citep[see, \emph{e.g.,} discussion in][]{Agertz+21}.
In principle, a clean kinematic separation between
thin and thick discs could be a natural way to interpret 
the stellar phase space densities, with ``high-density'' 
stars being thin disc members and ``low-density'' stars 
being thick disc members. However, when we use the 
rough kinematic classifications from \cite{Bensby+14}---thin 
disc stars at $|\mathbf{v}| < 50\mathrm{\,km\,s^{-1}}$ and 
thick disc stars at 
$70\mathrm{\,km\,s^{-1}}\lesssim |\mathbf{v}| \lesssim 
180\mathrm{\,km\,s^{-1}}$---we see that the ``low-density'' 
stars are drawn from both the thick disc and the 
heated end of the thin disc, with a handful 
of halo stars as well (see Figure~\ref{fig:V_rho}). 
It is likely that the thick disc stars formed on kinematically 
hot orbits early in the Galaxy's history due to early mergers 
and the turbluent nature of the gas disc at early times 
\citep{Bird+13,Agertz+21,Renaud+20}, and they have maintained 
their higher velocity dispersion \cite[e.g.,][]{Miglio+21} to 
the present day. This does not affect our argument, since these stars 
are all old and kinematically hot, and thus have a low phase 
space density.

As the ``high-density'' stars are kinematically cold and 
therefore on average young, and the ``low-density'' stars 
are a mix of old thick disc stars and old heated thin 
disc stars, the stellar age naturally suggests itself 
as an explanation for the overabundance of Hot Jupiters 
orbiting ``high-density'' hosts. Hot Jupiters can spiral 
in to their host stars under tidal drag 
\citep[\emph{e.g.,}][]{Jackson+09,Levrard+09,CCJ18}, and if the 
tidal dissipation is effective enough, the timescale for this
is also $\lesssim$\,Gyrs, similar to that for kinematic heating 
of the host star. \cite{HamerSchlaufman19} previously 
identified that Hot Jupiter hosts have colder kinematics 
than Cold Jupiter hosts, with similar values of the velocity 
dispersions to those we have found, and found that 
this difference corresponds to tidal decay of Hot Jupiter orbits 
if the tidal quality factor $Q_\star^\prime\lesssim10^7$. 
Hot Jupiter hosts, then, are predominantly in high-density
regions of phase space because of a bias towards detecting the 
Hot Jupiters around young stars before their tidal destruction, 
a bias noted by \cite{CCJ18}. A potential confounding
factor we have not considered is stellar metallicity, which
has a strong influence on the probability of forming
a giant planet \citep{Fischer05}. However, in the Solar neighbourhood
the age--metallicity relation is rather flat back to
ages of around $10$\,Gyr \citep{Freeman02,Sahlholdt21};
for very old stars such as the thick disc, both
metallicity and $\alpha$-element abundance may
affect planet formation \citep{Adibekyan+21b}.

We note that we have not shown that there is no 
impact of birth environment on planetary system 
architecture, only that the differences found 
through the phase space method 
primarily arise as a result of age and that nothing 
is seen when this confounding factor is removed, at 
least for the Hot Jupiters. Surveys directly looking 
at planets in clusters \citep[\emph{e.g.}][]{Rizzuto+20,
  Nardiello+20} could address this, but conclusions 
may be tentative because of the low yield of 
discoveries: \cite{Brucalassi+16,Brucalassi+17}
found a Hot Jupiter rate higher in the 
M67 open cluster than in the field,
but this relies on only three Hot Jupiters found in the cluster.
There are also other trends found by W20 and subsequent 
papers \citep{Kruijssen+20,Chevance+21,Longmore+21} that 
must be explained; we note however that the finding 
of \cite{Chevance+21} that there is a stronger gradient 
in planetary radius in multi-planet systems orbiting 
``low-density'' hosts may also be an age effect, as this 
gradient can result from photoevaporation of the 
planets' atmospheres \citep{OwenWu13} and the older ``low-density'' 
stars have more time for this process to proceed.
The trend in multiplicity found by \cite{Longmore+21}
is harder to explain: they found that low-density hosts have
more multiple systems than high-density hosts. This would seem to
go against an age dependence (multiplicity should reduce with
time), but we note that their good-quality low-density samples
had only 5 or 6 stars, and a larger sample would be required
to confirm or refute this.

We have followed W20 in using the heterogeneous 
sample of exoplanets provided by the NASA Exoplanet Archive. 
Recently, \cite{Adibekyan+21} 
used a smaller homogeneous sample to look for differences between 
the ``high-density'' and ``low-density'' populations; the sample
was unfortunately too small to see a significant difference. 
A difference did emerge in a larger sample, although 
\cite{Adibekyan+21} noted that the ``low-density'' hosts 
are older (with a homogeneous age determination) 
than the ``low-density'' hosts. This again 
underlines the importance of correcting for age. We finish with 
two further caveats for further studies that may wish 
to look for trends after the age dependence is removed: 
first, the differential completeness of \emph{Gaia} across 
a target star's neighbourhood should be accounted for 
(see Figure~\ref{fig:complete}); and second, 
coherent structures can arise in velocity space among 
stars of divers ages through interactions with 
matter inhomogeneities in the Galaxy 
\citep[\emph{e.g.,}][]{DeSimone+04,Antoja+18,Kushniruk+20}, so 
phase space overdensities need not reflect a coeval origin 
in a dense environment.\footnote{While                                                                  
  this paper was under review, \cite{Kruijssen21} submitted a paper
  linking the phase space densities to such features of 
  Galactic dynamics: the ripples within the Galactic disc
  generated by matter inhomogeneities such as the bar, arms and
  satellite galaxies.}

\section{Conclusions}

\begin{enumerate}
\item Classifying stars according to 
  their local 6D phase space densities, we 
  verify that Hot Jupiter hosts preferentially 
  belong to the population of high phase space density.
\item Phase space density shows an extremely 
  strong anti-correlation with a star's peculiar 
  velocity with respect to the local standard of rest 
  in the Galaxy. The high phase space density of Hot Jupiter 
  hosts is primarily a manifestation of their cold 
  kinematics.
\item After correcting for the dependency of phase space 
  density on peculiar motion, there is no evidence that 
  Hot Jupiter hosts lie in denser regions of phase 
  space than other stars.
\item The observed correlation is likely to 
  arise from the bias towards detecting Hot Jupiters 
  around younger (and therefore kinematically colder) 
  host stars, before the Hot Jupiters are destroyed by 
  tidal orbital decay.
\end{enumerate}
A Jupyter notebook and ancillary files 
to reproduce these results are
available\footnote{\url{https://github.com/AJMustill/HJGalaxy}}.

\begin{acknowledgements}
  AJM acknowledges funding from the Swedish Research 
  Council (grant 2017-04945), the Swedish National Space 
  Agency (grant 120/19C), and the Fund of the Walter 
  Gyllenberg Foundation of the Royal Physiographic 
  Society in Lund. This research has made use of 
  the Aurora cluster hosted at LUNARC at Lund University. 
  AJM wishes to thank Ross Church, Sofia 
  Feltzing, Diederik Kruijssen, Steve Longmore, Paul McMillan,
  Pete Wheatley, Andrew Winter and the anonymous referee for useful
  comments and discussions.
  This work has made use of data from the European Space Agency (ESA) mission
  {\it Gaia}\footnote{\url{https://www.cosmos.esa.int/gaia}}, processed by the {\it Gaia}
  Data Processing and Analysis Consortium (DPAC)
  \footnote{\url{https://www.cosmos.esa.int/web/gaia/dpac/consortium}}. Funding for the DPAC
  has been provided by national institutions, in particular the institutions
  participating in the {\it Gaia} Multilateral Agreement.
  This research made use of Astropy,\footnote{\url{http://www.astropy.org}} 
  a community-developed core Python package for 
  Astronomy \citep{astropy:2013, astropy:2018}. 
  This research made use of NumPy \citep{2020NumPy-Array}, 
  SciPy \citep{2020SciPy-NMeth}, MatPlotLib \citep{2007CSE.....9...90H},
  and Scikit-learn \citep{scikit-learn}. 
  This research has made use of the NASA Exoplanet Archive, 
  which is operated by the California Institute of Technology, 
  under contract with the National Aeronautics and Space Administration 
  under the Exoplanet Exploration Program.
\end{acknowledgements}

%
\bibliographystyle{aa} 
\bibliography{HJGalaxy.bib} 
%

\appendix

\section{Additional figures}

\subsection{Colour--magnitude diagram for Solar neighbours}

\begin{figure*}
  \centering
  \includegraphics[width=0.49\textwidth]{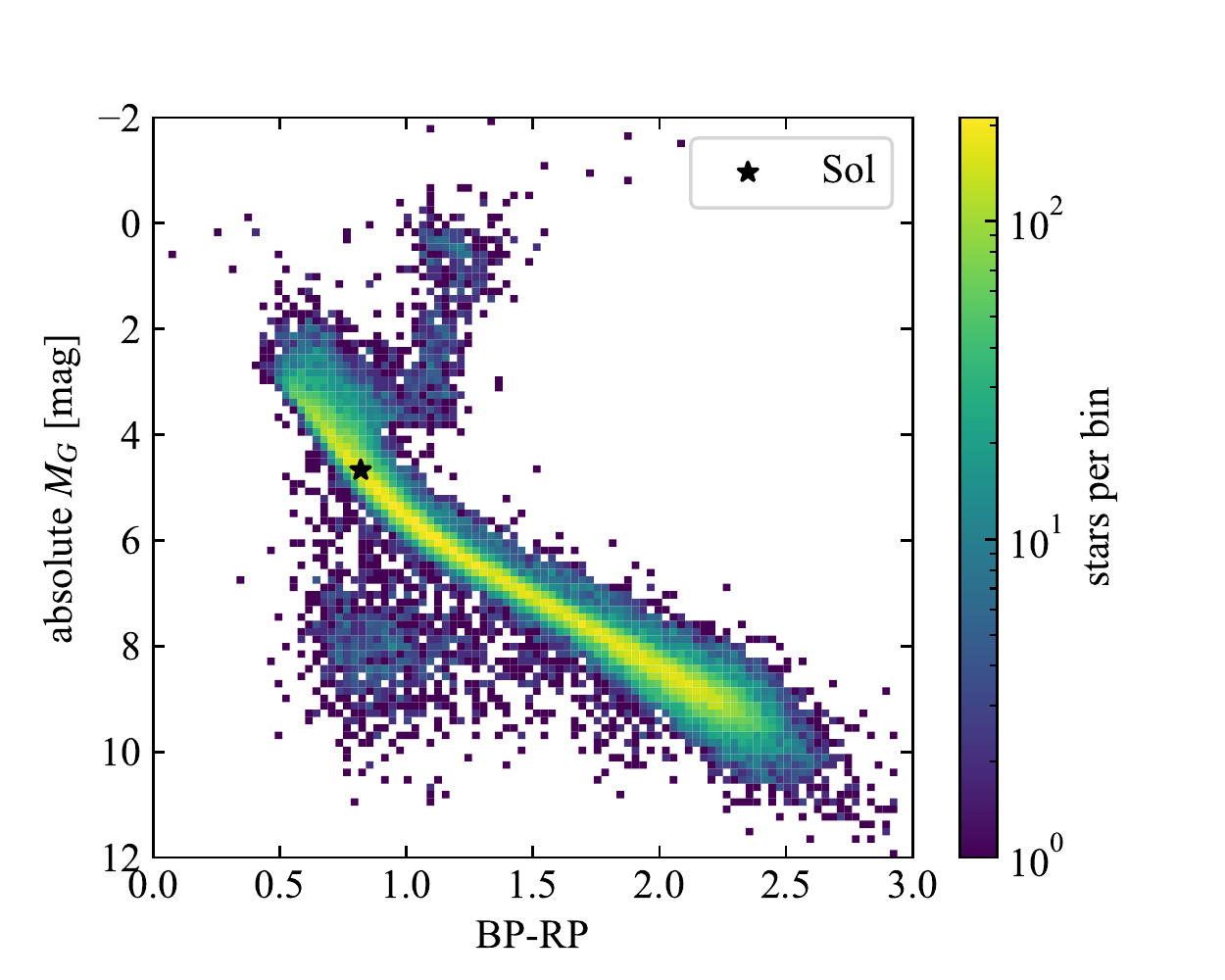}
  \includegraphics[width=0.49\textwidth]{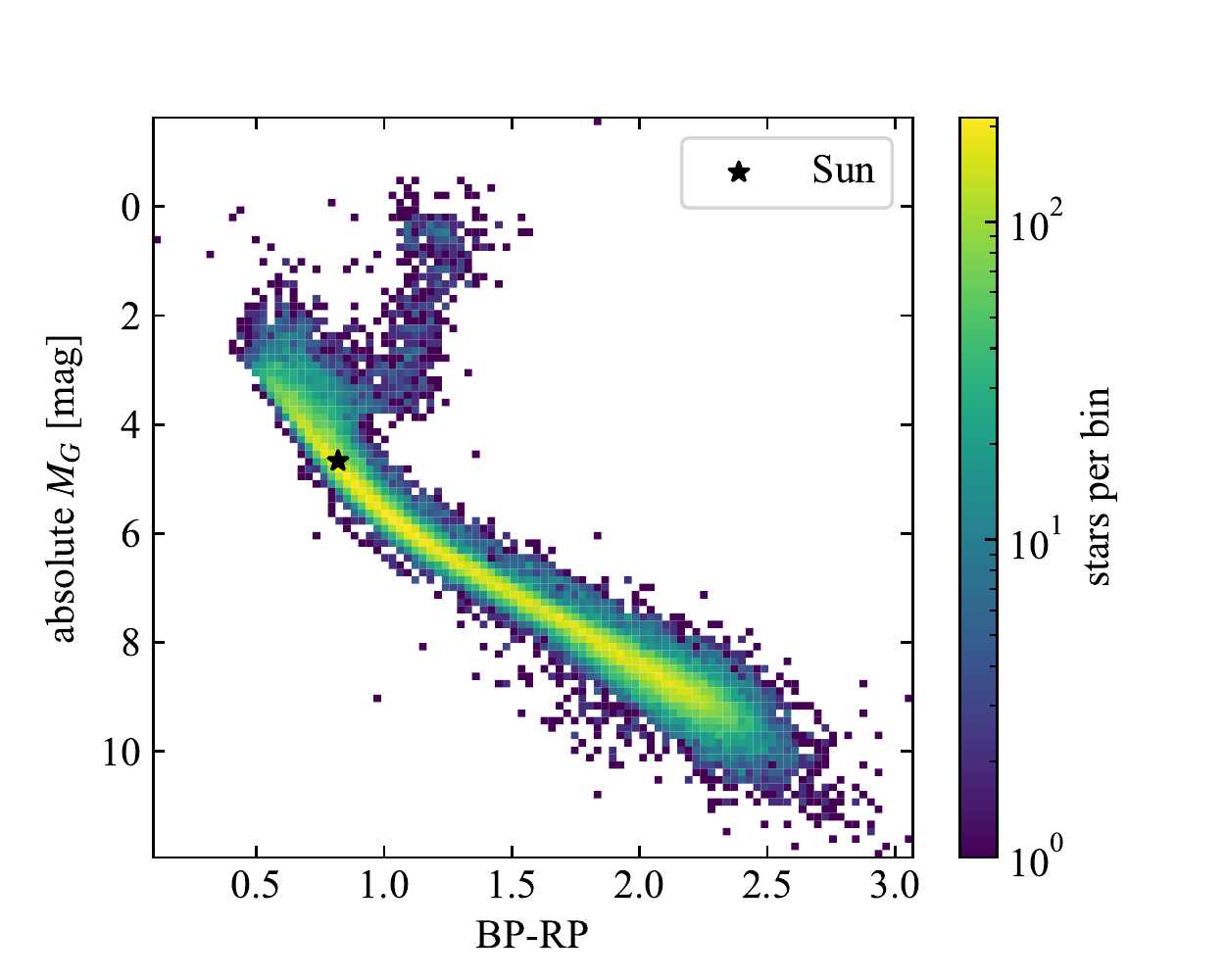}
  \caption{Colour--magnitude diagrams for all stars within 80\,pc
    of the Sun with \emph{Gaia} RVs. Left: DR2. Right: EDR3.
    The \emph{Gaia} colour and magnitude for the Sun are
    taken from \cite{Casagrande+18}.}
  \label{fig:CMD}
\end{figure*}

We show the colour--magnitude diagram for our sample
of Solar neighbours in Figure~\ref{fig:CMD}. We have, in keeping with W20, imposed
no quality cuts on the \emph{Gaia} data (other than the need for
parallax to be positive); however, note that moving from DR2 to EDR3
cleans the CMD considerably, especially of spurious objects below
the main sequence. The blob at $M_G\sim8$, below the main sequence, 
is discussed in \cite{GaiaNearby20}, who attributed it to a change 
at $G=13$ in the window of pixels on the CCD surrounding around the target 
\citep{Carrasco+16}.  
In this paper, we work with the EDR3 data.

\subsection{Gaussian mixture model}

\begin{figure*}
  \centering
  \includegraphics[width=\textwidth]{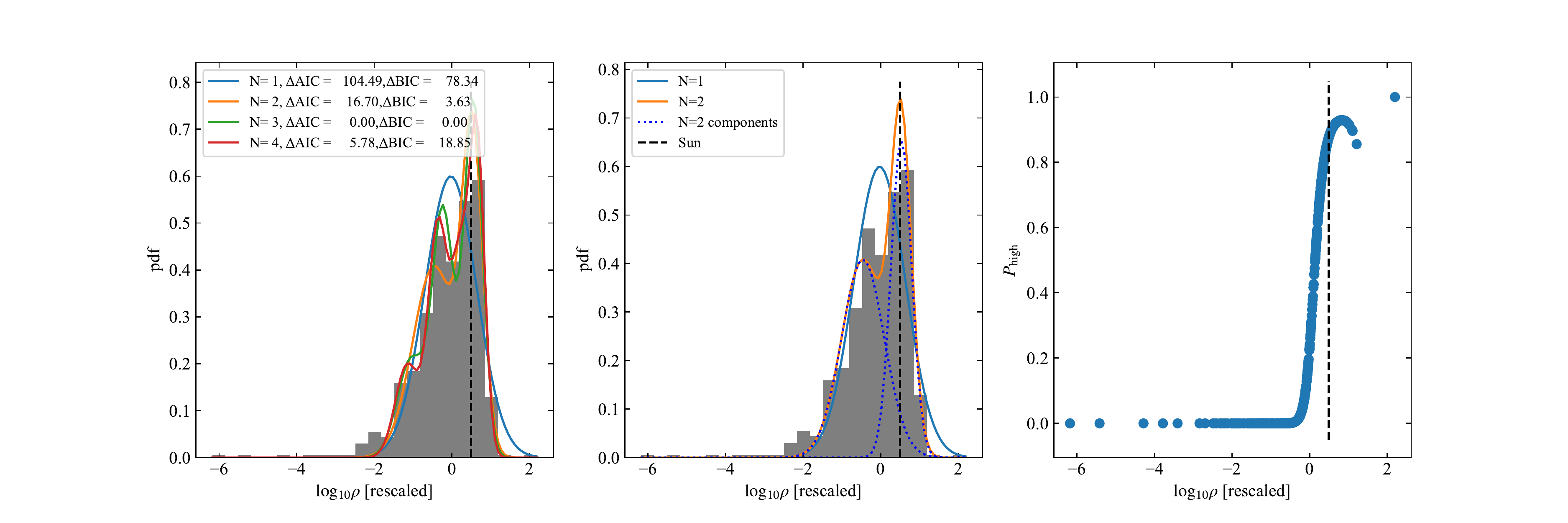}
  \caption{Gaussian Mixture models for the
    local phase space density of the Sun. Left: one- to ten-component
    models, together with Aikake and Bayesian Information
    criteria values relative to the best fit. Middle: The one-
    and two-component models, together with the decomposition of
    the latter. Right: probability each star belongs to the high-density
  component of the two-component model.}
  \label{fig:GMM}
\end{figure*}

\begin{figure*}
  \centering
  \includegraphics[width=\textwidth]{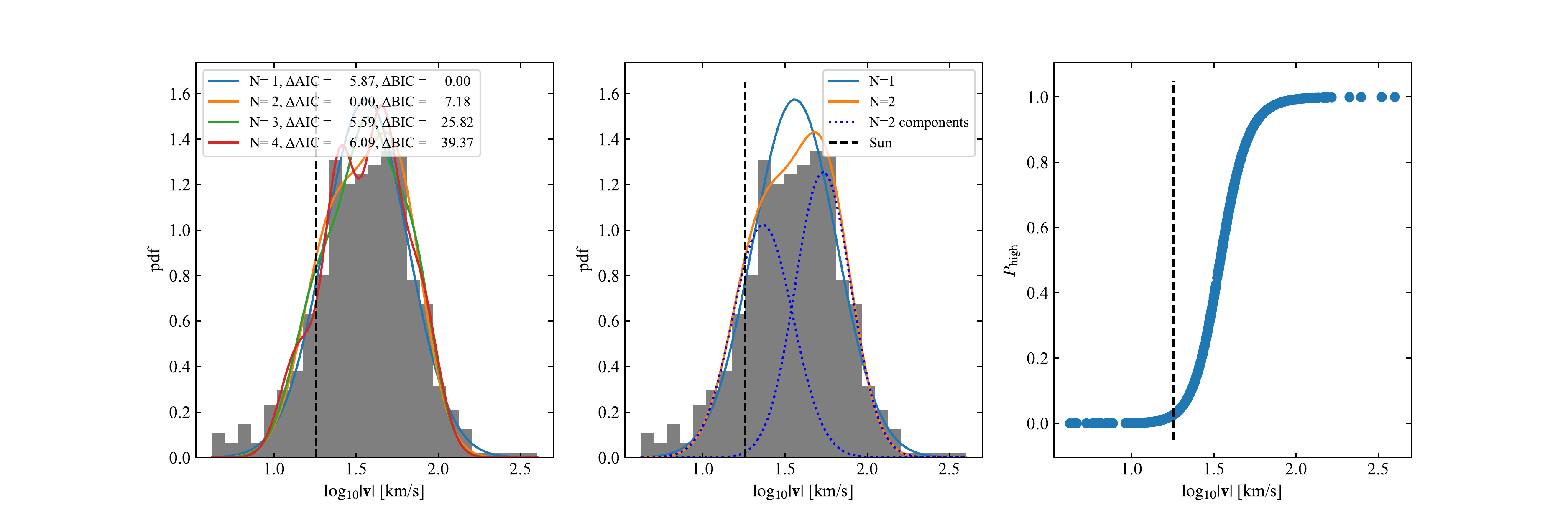}
  \caption{Gaussian Mixture models for the
    local peculiar velocity distribution of 
    neighbours of the Sun. Left: one- to ten-component
    models, together with Aikake and Bayesian Information
    criteria values relative to the best fit. Middle: The one-
    and two-component models, together with the decomposition of
    the latter. Right: probability each star belongs to the high-velocity
    component of the two-component model.}
  \label{fig:GMM_v}
\end{figure*}

In Figure~\ref{fig:GMM} we show the Gaussian mixture model 
for the Sun and 600 of its neighbours, together with the 
stars' probability of belonging to the high-density population.
Note that this probability begins to decrease slightly for 
stars with moderately high densities, on account of the 
breadth of the low-density distribution. Stars at $\rho>50$ 
are assigned $P_\mathrm{high}=1$ to alleviate this.
We show the equivalent decomposition in 
velocities in Figure~\ref{fig:GMM_v}.

\subsection{Spatial structure}

\begin{figure*}
  \centering
  \includegraphics[width=0.49\textwidth]{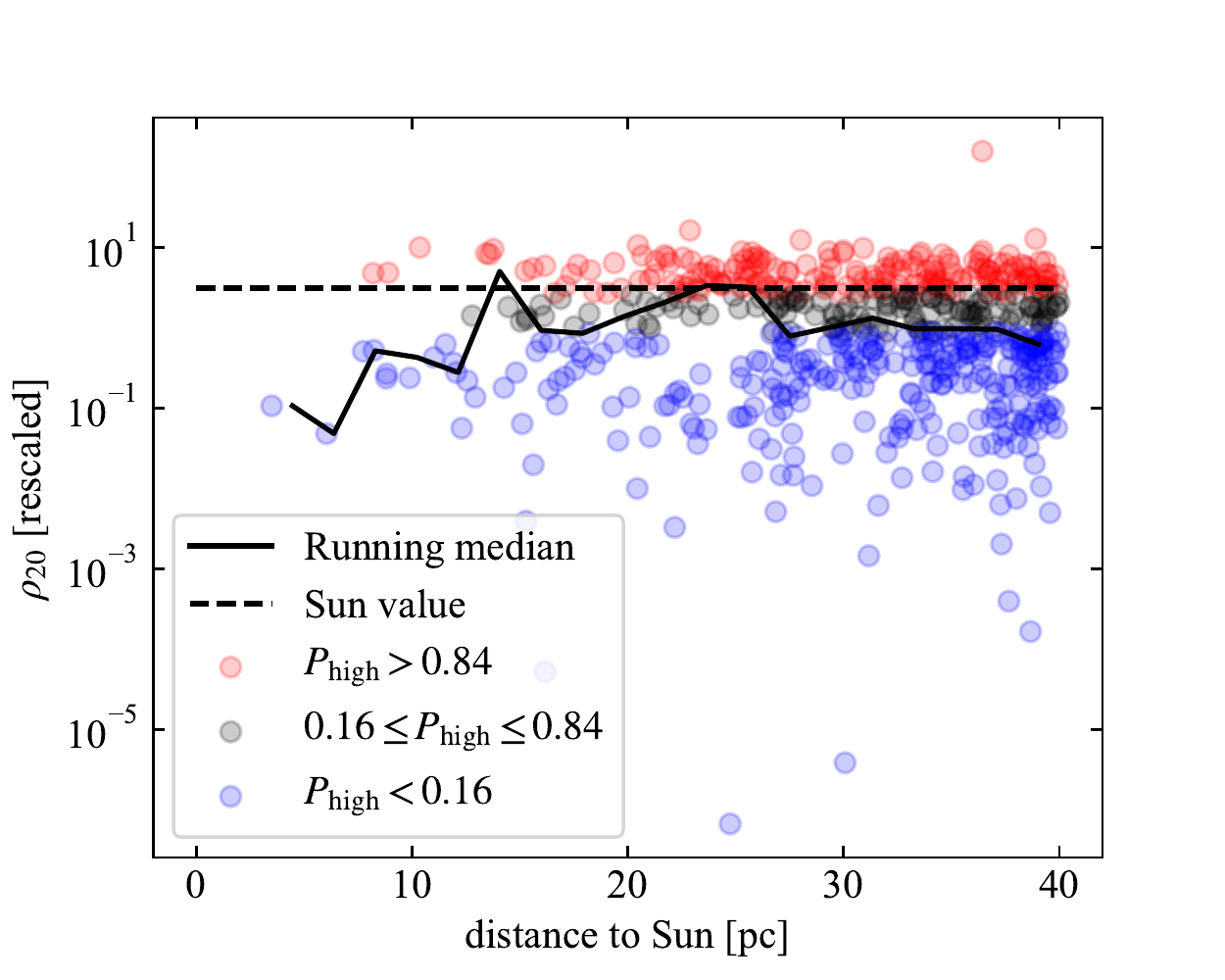}
  \includegraphics[width=0.49\textwidth]{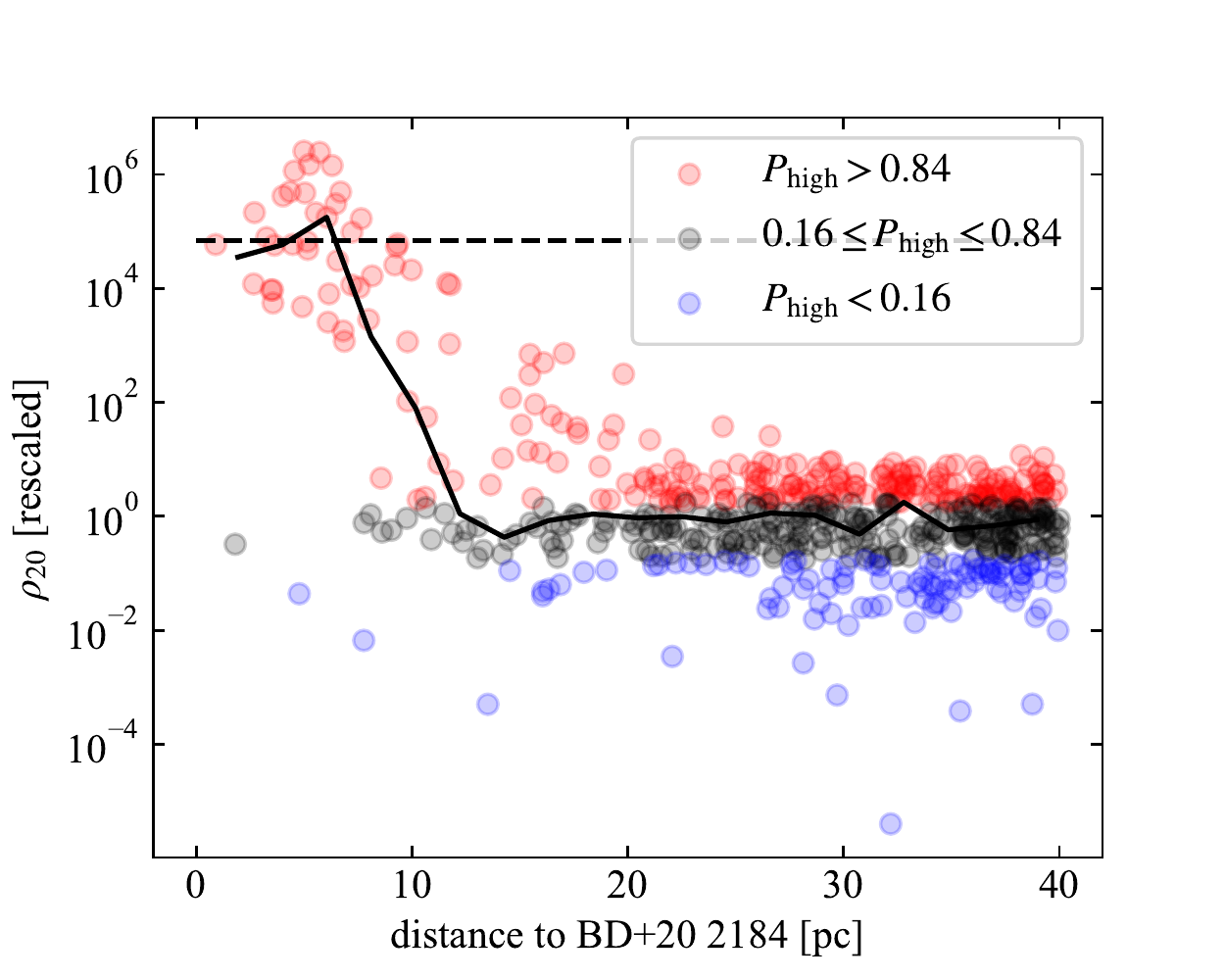}
  \caption{Local 20$^\mathrm{th}$-nearest neighbour
    phase space density, and probability
    of membership in the high-density population,
    for 600 stars within 40\,pc of the Sun (left)
    and Pr0201 (BD+20~2184, right). We show these
    as a function of the distance to the target star.}
  \label{fig:D_rho}
\end{figure*}

In Figure~\ref{fig:D_rho} we show the phase space density 
as a function of distance to the target star, for 600 neighbours of 
the Sun and of the Pr{\ae}sepe member BD+20~2184. There is little structure 
seen among the neighbours of the Sun, although there may be 
a weak trend from 25 to 40\,pc; this disappears if we restrict 
attention to stars with an absolute magnitude $M_G\le8$ and may 
reflect the decreasing completeness of \emph{Gaia} with distance. 
For BD+20~2184, fellow cluster members 
stand out at high density close to the target.

\begin{figure*}
  \centering
  \includegraphics[width=0.8\textwidth]{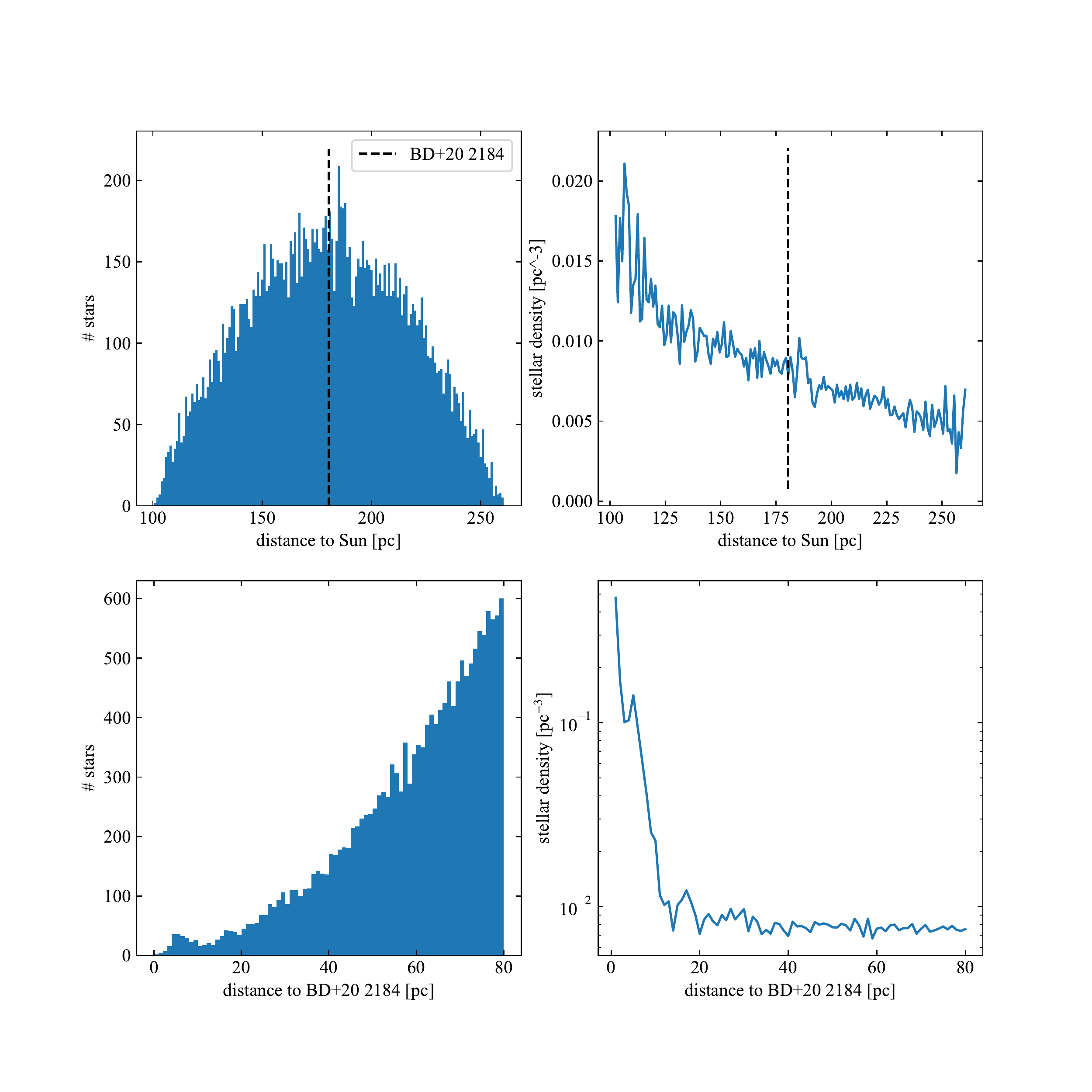}
  \caption{Differential completeness of the \emph{Gaia}~EDR3 RV 
    catalogue across a sphere of 80\,pc centred on BD+20~2184. 
    Left-hand panels show histograms of the number of stars per distance 
    bin, measured from the Sun (top) and from BD+20~2184 (bottom). 
    Right-hand panels show the corresponding 3D spatial density.}
  \label{fig:complete}
\end{figure*}

We show in Figure~\ref{fig:complete} the differential completeness 
of the \emph{Gaia}~EDR3 RV
catalogue across a sphere of 80\,pc centred on BD+20~2184.
Looking at the top panels, we see that the 3D spatial density 
of stars in the RV catalogue falls by a factor of three 
between the near side of the sphere and the far side of the 
sphere. This seems not to induce a strong bias in the 
spatial density as a function of distance from 
BD+20~2184 itself, shown in the bottom right panel: 
the Pr{\ae}sepe cluster clearly stands out 
as a significant spatial overdensity, beyond 
which the density is constant. Nevertheless, the 
potential for the differential completeness to 
induce a bias should be borne in mind as a potential confounding 
factor once the main velocity dependence of the phase 
space density is accounted for.

\subsection{Alternative detrending comparison}

\label{sec:compare}

\begin{figure*}
  \centering
  \includegraphics[width=0.33\textwidth]{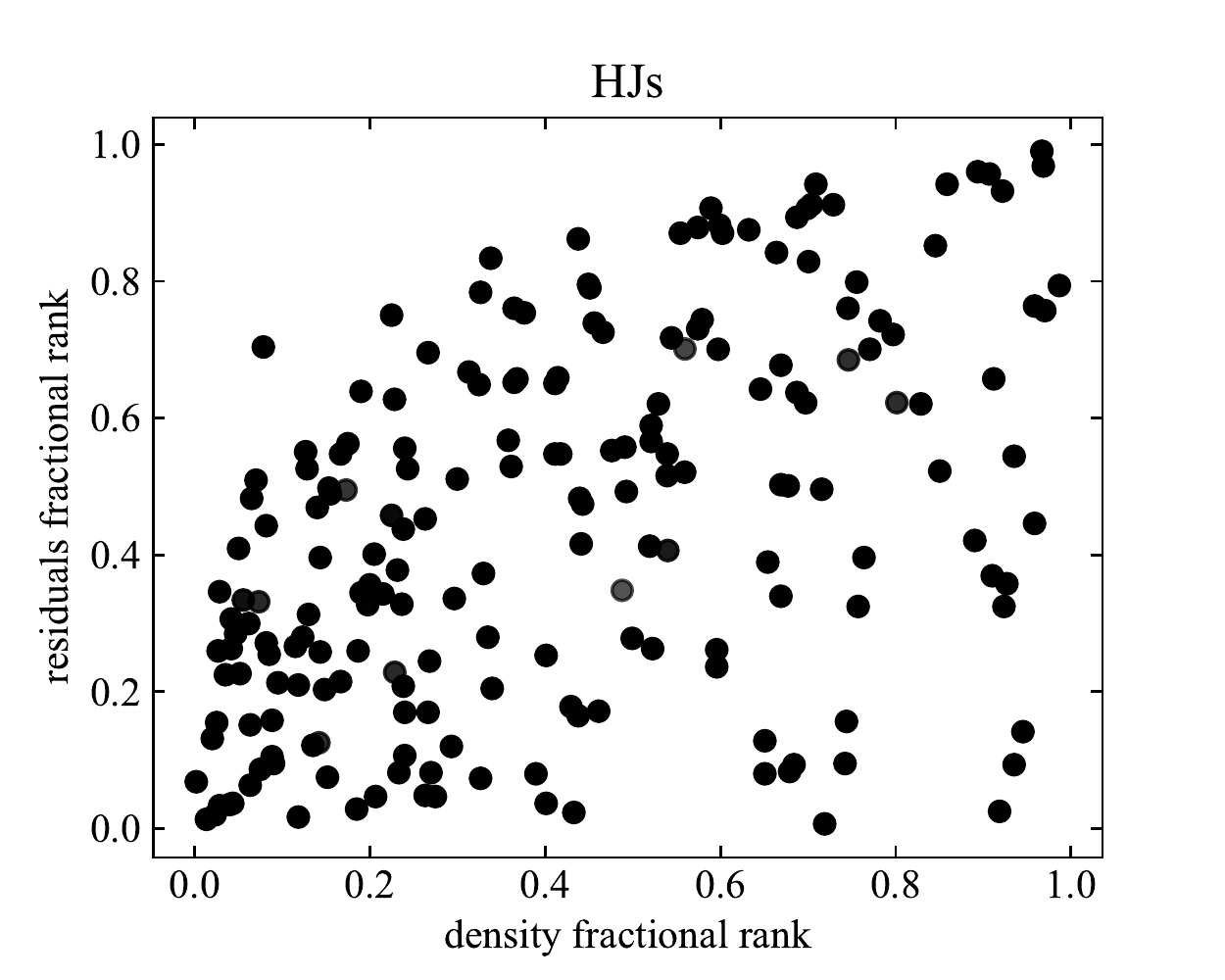}
  \includegraphics[width=0.33\textwidth]{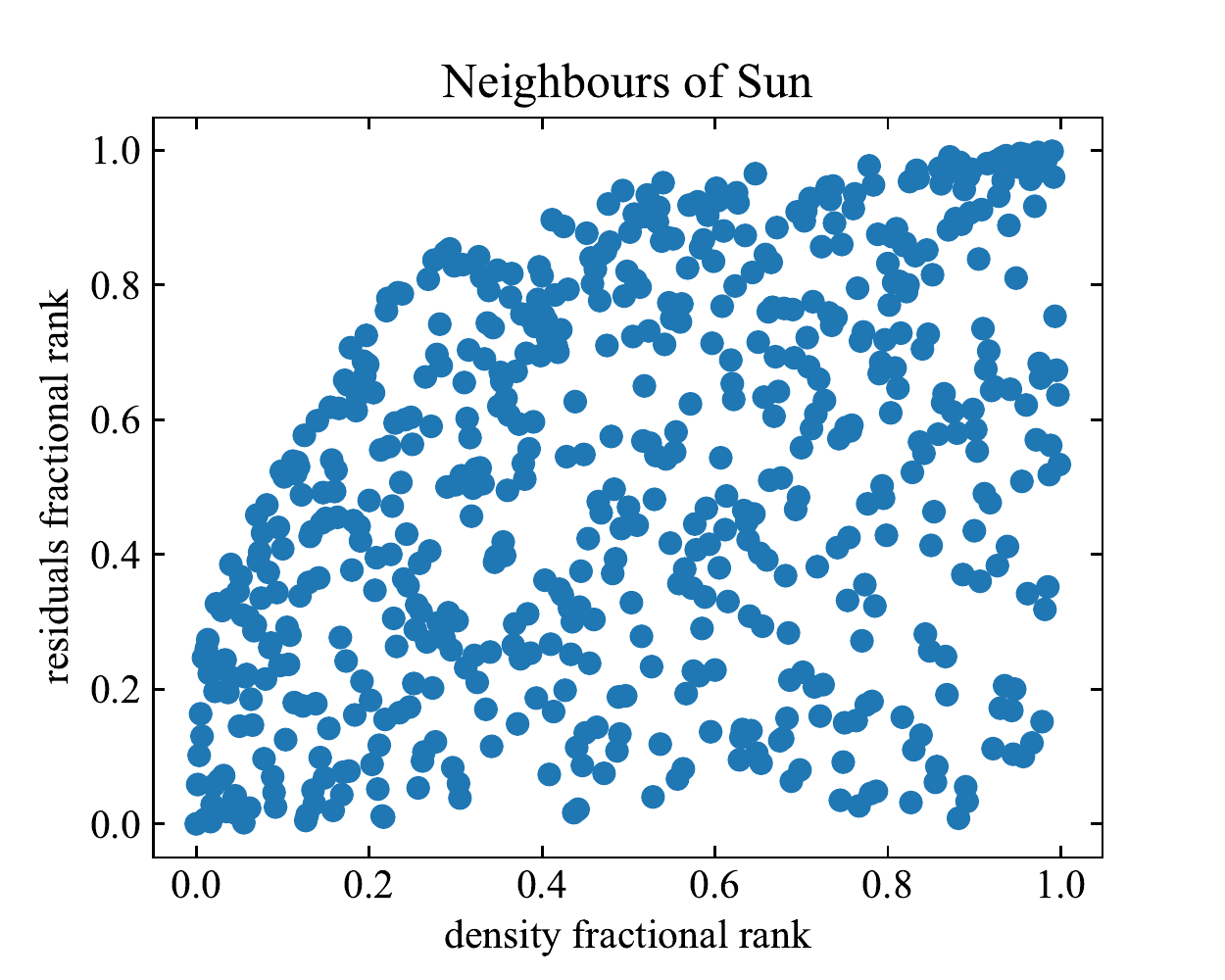}
  \includegraphics[width=0.33\textwidth]{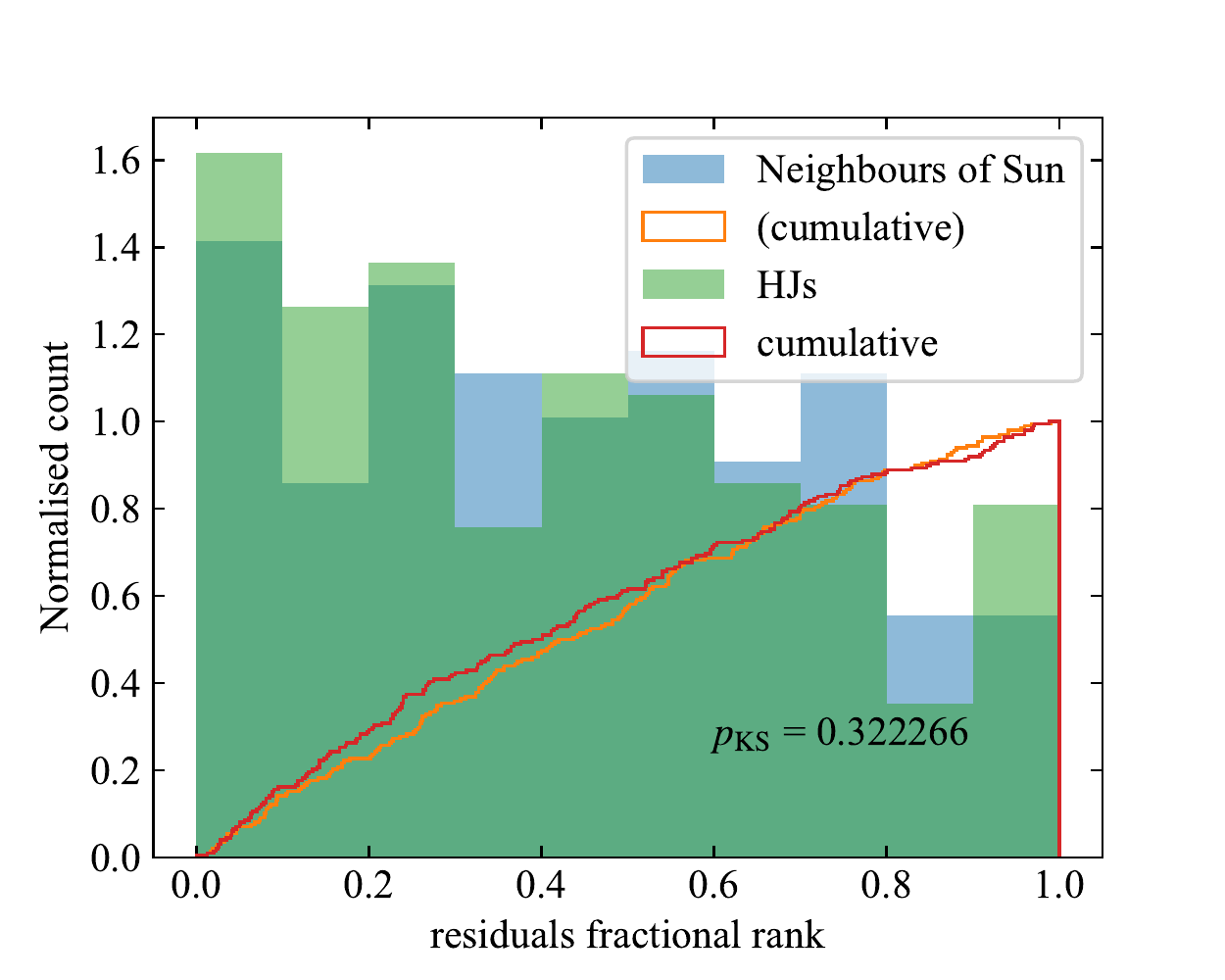}
  \caption{Left: The fractional rank (rank among neighbours divided by
    number of neighbours) in phase space density, and in residuals
    to the detrended phase space density, for all the 198 Hot Jupiter
    hosts. Darker symbols are stars with more neighbours in
    \emph{Gaia}~EDR3. Centre: The same, for 600 neighbours
    of the Sun. Right: The histogram of the fractional residuals
    rank for the Hot Jupiter hosts (blue for differential,
    orange for cumulative), and for the fractional residuals
    rank of 198 neighbours of the Sun chosen to have the same
    density fractional rank as the Hot Jupiter hosts (green
    for differential, red for cumulative). The KS test $p$-value
    comparing the two distributions is also displayed, 
    showing that they are statistically indistinguishable.}
  \label{fig:ranks}
\end{figure*}

In the main body of the paper (\S\ref{sec:residuals}), 
we showed that the phase space density residuals, after 
the velocity trend is removed, are identical for 
the Hot Jupiter hosts and for the Cold Jupiter hosts. 
This implicitly puts all of the phase space densities on 
the same scale, despite their being constructed from 
different samples. While the densities are all normalised 
such that the median of each distribution is unity, 
and the fitted trends are all quite similar, nevertheless 
there is a small scatter in the fitted trends. Here we 
descrbe an alternative test that relies only on the ranks 
of stars within their populations of neighbours, 
both in terms of phase space density and 
the residuals after detrending.

As each sample may contain a different number of stars, 
we first define fractional ranks between 0 and 1 such that
\begin{equation}
f_\rho = r_\rho/N_\mathrm{sample}
\end{equation}
and
\begin{equation}
f_\mathrm{res} = r_\mathrm{res}/N_\mathrm{sample},
\end{equation}
where $r_\rho$ and $r_\mathrm{res}$ are the ranks 
(1 being high) of each star in density and in the residuals, 
and $N_\mathrm{sample}$ is the number of stars in the sample.
We now compare the Hot Jupiter hosts to the neighbours of the Sun. 
For each Hot Jupiter host, we calculate its fractional ranks, 
and pair it up with the nearest-ranked neighbour of the Sun, with 
rank (in density)
\begin{equation}
  r_\mathrm{comp} = \mathrm{floor}\left(N_\mathrm{control} f_{\rho\mathrm{,HJ}} \right),
\end{equation}
where $N_\mathrm{control}=600$ is the number of neighbours of the Sun, 
and $f_{\rho\mathrm{,HJ}}$ is the fractional density rank of the 
Hot Jupiter host. We then construct a control distribution 
of residuals by selecting the fractional residuals ranks
of the Solar neighbours whose density ranks are given by 
$r_\mathrm{comp}$.

This procedure is illustrated in Figure~\ref{fig:ranks}.
The fractional ranks for the Hot Jupiter hosts are shown
in the left-hand panel. The density of points
increases towards the bottom left. The increase towards the left
reflects the tendency for Hot Jupiter hosts to have high
phase space densities. The increase towards the bottom shows that they tend
to have high residuals after detrending. 
However, even an unbiased
sample can show a tendency for high density to correspond with
high residuals. This is demonstrated in the middle panel of
Figure~\ref{fig:ranks}, where we show the equivalent rankings for
the 600 neighbours of the Sun. 
In the right-hand panel of Figure~\ref{fig:ranks}, 
we show the comparison between the Hot Jupiter 
sample and the control sample drawn from 
the centre panel. The Hot Jupiter sample and the control sample
are then indistinguishable on a KS test ($p=0.32$). Note that a set
of randomly-selected neighbours of the Sun would give a uniform
distribution.

\end{document}